# Enhancing Financial Decision-Making: Machine Learning and AI-Powered Predictions and Analysis.

Vishal Patil, Kavya Bhand, Kaustubh Mukdam, Kavya Sharma, Manas Kawtikwar, Prajwal Kavhar, Hridayansh Kaware

*Abstract - The proposed system aims to use various machine learning algorithms to enhance financial prediction and generate highly accurate analyses. It introduces an AI-driven platform which offers inflation-analysis, stock market prediction, and E-learning module powered by a chatbot. It has achieved high accuracy where the Inflation Analysis depicts 0.8% MAE, 1.2% RMSE and the Stock Prediction shows 98% and 96% accuracy for Apple and Google stock prices respectively. Key features include historical price trends, inflation rates, short-term future stock prediction, where the data has been extracted using real-world financial datasets. Additionally, the E-learning feature contributes to bridging financial gaps and promoting informed decisions. We have implemented algorithms like linear regression, ARIMA, LSTM where the accuracy has been evaluated using metrics such as MAE, RMSE and the like.*

I. INTRODUCTION

Finance is the backbone of any economy which guides both individual as well as collective prosperity. At every level, finance has significant consequences on how the resources are utilized, how one manages the potential risks and how one creates and distributes wealth. In today's world, financial literacy and access to precise and accurate financial tools have become a critical component of economic stability. However, a significant portion of the population, mainly underserved communities, still face barriers in gaining financial education, resources, and inclusive financial services.[4]. This paper proposes to bridge these gaps using advanced techniques such as data analytics, artificial intelligence, and machine learning. This project's idea is to design an integrated platform that can help provide users with instruments for economic trends analysis, forecasting of the movement of the stock market, controlling the risks of investments, and providing interactive modules for e-learning on these topics. This project aims at empowering individuals and organizations with sufficient knowledge to take informed decisions about financial issues and enhance their learning skills through interactive learning modules. The ML algorithms demonstrate possible potential in more informed financial decision making by achieving more accurate, efficient and promising results.[3]. By having access to comprehensive education through the e-learning modules, it will serve as an advantage to the users for individual development and social good.

II. LITERATURE REVIEW

The fast-paced advancements in AI and ML have transformed financial analytics. Multiple studies demonstrate the efficient use of machine learning in financial forecasting which include inflation rate prediction, stock market trend analysis and risk management to be precise. Traditional methods often rely on statistical techniques ultimately struggling to capture the non-linear patterns in a large dataset. Deep learning models such as Long Short-Term Memory (LSTM) networks have gained immense value because of their ability to predict market trends with high accuracy while handling sequential financial data.

Researchers have also dived into the idea of integrating AI-driven chatbots to address finance based topics. These assist in enhancing financial literacy by providing real-time recommendations and personalised financial guidance. Past research [2, 3] has highlighted the effectiveness of ARIMA and LSTM models in inflation and stock prediction and the results produced surpassed conventional econometric models. This proposed system builds upon the existing research by utilising various ML techniques in an all-inclusive AI-powered financial prediction and learning platform.

## III. METHODOLOGY/EXPERIMENTAL

### 1. Inflation Analysis

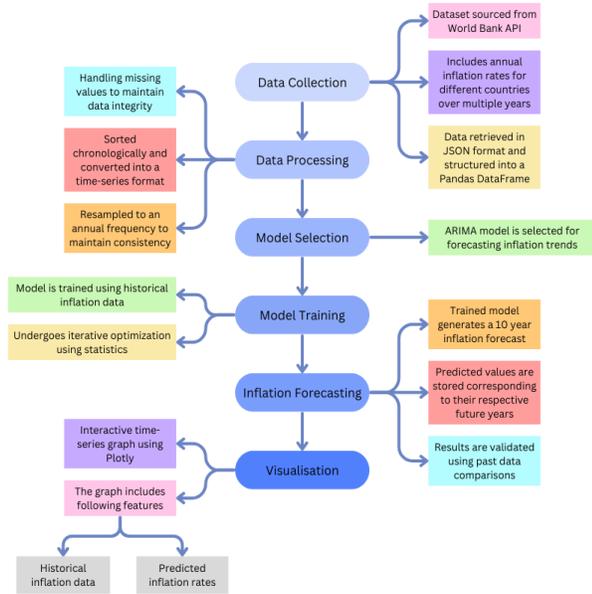

Fig. 1. Flowchart for the Inflation Analysis model.

#### 1.1. Data Acquisition & Preprocessing

The dataset is fetched from the World Bank API ranging from the earliest year data that is available for the selected country. This data is converted into a Panda DataFrame followed by removing the missing values and indexing using datetime format. This is then sorted chronologically and resampled to annual frequency (YS) to ensure data integrity and consistency in prediction.

#### 1.2. Time Series Modeling and Forecasting

An ARIMA (p, d, q) model is selected to forecast inflation trends. These parameters train the model and generate a 10-year forecast.
p = 15, where it uses inflation rates from the past 15 years
d = 1, which is the differencing applied for stationarity
q = 0, that is no moving average

#### 1.3. Interactive Data Visualisation

The trends are visualised using Plotly where the historical inflation is depicted using blue solid line and predicted inflation is depicted using red dashed line. Interactive features include hover-enabled data points, pan, zoom, which enhances user engagement.

### 2. Stock Prediction

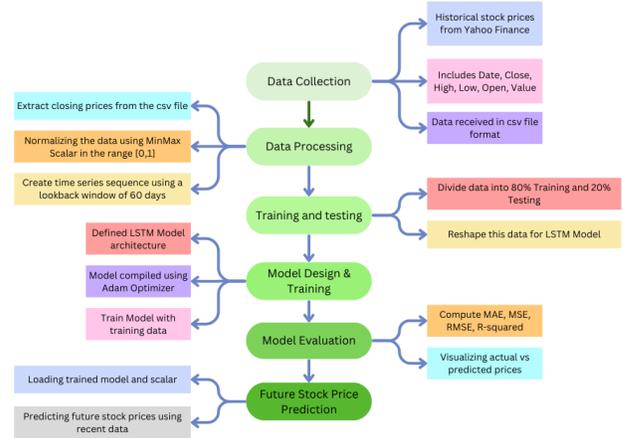

Fig. 2. Flowchart for the Stock Prediction model

#### 2.1. Data Collection

Historical stock market data is obtained from Yahoo Finance. The dataset includes daily stock prices from early 2000 till the latest day. The closing price is used as the primary feature for prediction.

#### 2.2. Data Preprocessing

To prepare the dataset for model training, we had extracted only the 'Close' price column from the CSV file generated from Yahoo Finance. This close price data was then normalized in the range of 0 to 1 using the MinMaxScalar to improve the efficiency of the model. A time series was constructed using a 60-day lookback window, where each sequence contains the past 60 days of price data as input and predicts the following day's price as output.

#### 2.3. Training and testing split

The preprocessed data is then split into two sets which are training (80%) and testing (20%) sets. The sequences are then reshaped according to the required format by the Long Short-Term Memory (LSTM) neural networks.

#### 2.4. Model Design & Training

The LSTM-based deep learning model is designed and trained in such a way that the architecture consists of two LSTM layers, each storing 50 units and a dropout layer so to prevent overfitting. Secondly, the model is combined with the help of mean squared error (MSE) loss function. And lastly, training of the model is performed with a batch size of 32 over 10 epochs using the training dataset.

#### 2.5. Model Evaluation

After training and testing, the model is then evaluated using several performance matrices which include Mean Absolute Error (MAE) which measures the average absolute difference between the actual and predicted stock prices, Mean Squared Error (MSE) which measures the average squared difference between actual and predicted values, Root Mean Squared Error (RMSE) which represents a more measurement of error magnitude as compared to MSE, and finally R-Squared ($R^2$) which measures how well the model explains the changes taking place in the stock price. Additionally, a graph is also plotted for showing the actual vs predicted prices for a better understanding of the model.

Fig. 3. Model summary for Apple Stock Prices (AAPL)

Fig. 4. Model summary for Google Stock Prices (GOOGL)

2.6. Future Stock Price Prediction

After training and evaluating the model, it is then used to forecast future stock prices. The trained model and the normalized close price data (using MinMaxScalar) are used for interference. The model takes last 60 days of closing prices of a stock as input to generate prediction for the next 5 days.

3. AI Chatbot

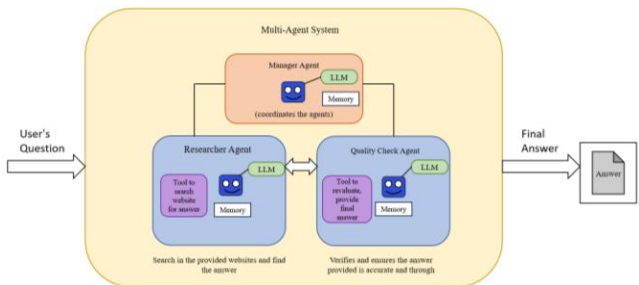

Fig. 5. Block diagram for AI Chatbot

## IV. RESULTS AND DISCUSSIONS

1. Inflation Analysis

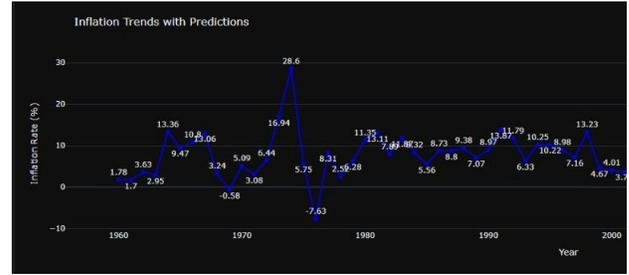

Fig. 6. Historical inflation trend for India

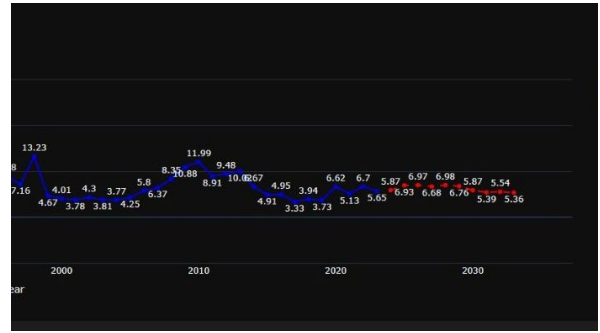

Fig. 7. Predicted inflation trend for India

Figure 6 and 7 displays the past and future inflation rates predicted for India. The blue solid line illustrates historical inflation trends and show significant fluctuations over the decades. Notable peaks can be observed during the 1970s and 1980s where the inflation rate went over 28.6% in 1974. In addition to that, the inflation rate experienced periods of sharp decline during the late 1970s. The future predictions depicted by red dotted line indicate a relatively consistent and stable inflation graph, averaging around 5.5% to 6.9% from 2025 to 2035. This suggests controlled economic fluctuations compared to highly fluctuated historical rates.

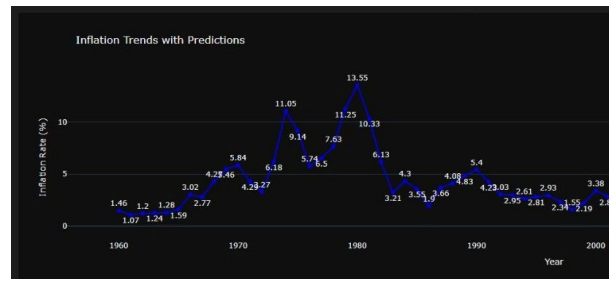

Fig. 8. Historical inflation trend for the United States

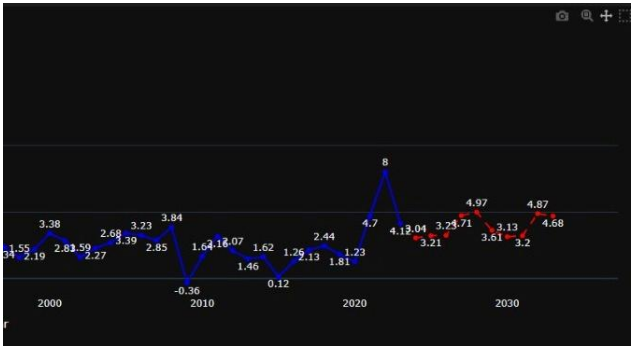

Fig. 9. Predicted inflation trend for the United States

Figure 8 and 9 illustrates the past and predicted inflation rates for the United States. The historical data depicts that the inflation rate peaked at 13.55% in the late 1970s and remained highly unpredictable until the 1990s after which it stabilized below 5%. The model predicts a stable inflation trajectory in the upcoming years with predicted values ranging between 3.1% and 4.9% from 2025 to 2035. Unlike India, the United States inflation pattern depicts a much controlled economic fluctuation, indicating a historically stronger monetary policy framework.

The trends are analysed for the United States and India. The model forecasts inflation rates for the next 10 years. The accuracy metrics are as follows :
  MAE - 0.8%
  RMSE - 1.2%
The results are visualised through interactive graphs after implementing ARIMA.

  2. Stock Prediction Module

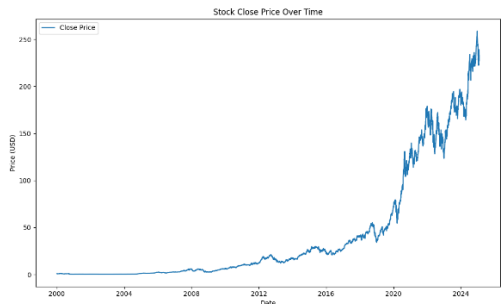

Fig. 10. Historical stock price graph for Apple (AAPL)

```
Mean Absolute Error (MAE): 4.33
Mean Squared Error (MSE): 30.40
Root Mean Squared Error (RMSE): 5.51
R-squared (R²): 0.98
```

Fig. 11. Accuracy metrics for AAPL

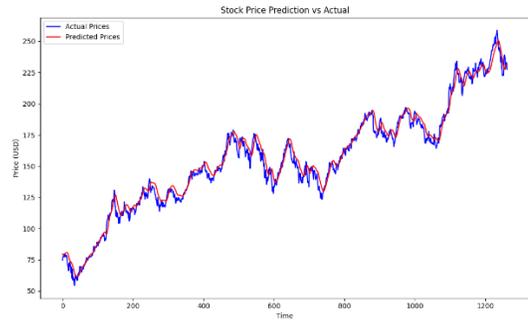

Fig. 12. Actual v/s Predicted stock price graph for AAPL

```
Predicted Future Stock Prices:
Day 1: $229.47
Day 2: $387.51
Day 3: $491.11
Day 4: $549.63
Day 5: $583.51
```

Fig. 13. Predicted future stock prices for AAPL

In the figure 10, historical stock price was analyzed where the data which included date, close, open, high, low, and volume of a stock was present in a csv file and a graph was plotted on the same. Looking at the trend of the stock price, it has shown a strong upward trend over time with some volatility and rapid growth. The stock price has shown a significant increasing trend from 2016, indicating a strong positive movement in stock price.

This model demonstrated a high accuracy $R^2$ score of 0.98 which explains that the model can explain 98% of variance in stock prices. The error matrices also supports the model accuracy with MAE coming around 4.33, MSE around 30.40, and RMSE around 5.51 which shows that model's prediction closely follows the actual stock price. The same values are shown in figure 11.

In figure 12, comparison between actual and predicted stock price after training and evaluating the model are shown in the form of graphs where blue line represents actual stock price and red line represent predicted stock price. After going through the graph, it is clear that the model efficiency captures stock price movements.

This model after evaluating the historical stock price gave prediction of future stock prices considering a window of 60 days of previous stock price and generating five days of future stock price. The values of the same are shown in figure 13.

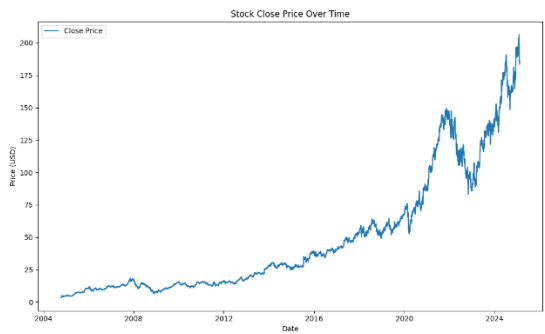

Fig. 14.  Historical stock price graph for Google (GOOGL)

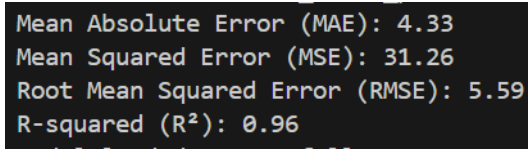

Fig. 15.  Accuracy metrics for GOOGL

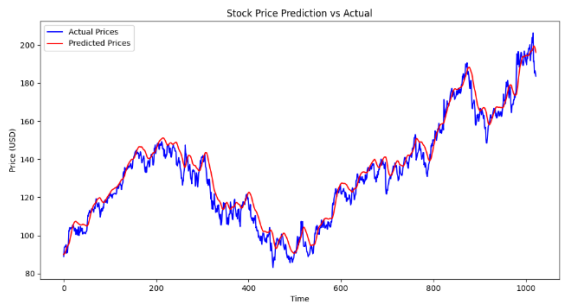

Fig. 16.  Actual v/s Predicted stock price graph for GOOGL

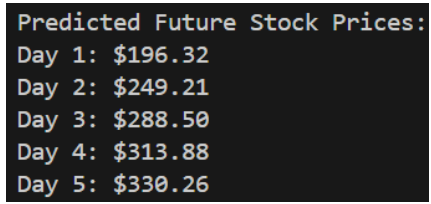

Fig. 17.  Predicted future stock price for GOOGL

Figure 14 shows the historical stock price graph for GOOGL where it shows the closing price of the stock on the y axis and the year/date on the x axis. The graph shows an upward trend with some fluctuations

In figure 15, the model shows the error metrices which consist of MSE being 31.26, MAE being 4.33, RMSE being 5.59, and $R^2$ being 0.96. This shows a higher $R^2$ value which indicates a well-fitted model that shows effective prediction of stock movement.

Comparison between actual and predicted stock price is shown in figure 16 in the form of graph where blue line shows the actual price of the graph whereas red line shows the predicted price of the graph after processing the historical stock data. After a close observation, it shows that the model has the ability to follow market trends with reasonable accuracy.

Based on model's training, evaluation and processing the previous stock data, it predicts future stock price of five days which suggests an overall positive trajectory. This prediction is based on the previous 60 days stock data and the prices for the next 5 days are shown in figure 17.

3. AI Chatbot

3.1. Multi-Agent Architecture:
Every agent has a predetermined role, goal and backstory which it incorporates to in order to deliver responses which are amiable, precise and have rich content.

3.2. Support Agent:
Its main responsibility is to answer user queries.
The goal of this agent is to offer the best user support.

3.3. Quality Assurance Agent:
The objective of this agent is to proofread and modify the answer given by support agent to make sure it is correct, detailed and polite. This agent conducts an additional check in order to guarantee all areas of the question have been addressed.

3.4. Query Solving Task:
Provides understanding of the particular user query with the user information. Makes use of all accessible tools to give an exhaustive answer.

3.5. Quality Inspection Task:
Assesses the support agent's answer. Guarantees that the end result satisfies the quality requirements.

3.6. Using Crew AI Framework:
This chatbot effortlessly handles and organises the operations of various agents and their respective responsibilities by means of Crew AI Framework. It facilitates agents working together to accomplish tasks.

3.7. Scrape Website Tool:
It retrieves content relevant to the user's query and provides credible data from external soured in order to improve the agent's response.

3.8. Using LLM:
The Groq LLM (groq/llma3-8b-8192), the engine behind both agents guarantees superior natural language processing and generation.

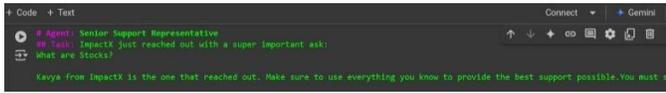

Fig. 18. User's question along with user details.

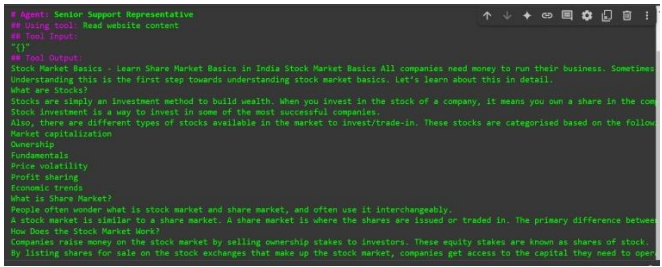

Fig. 19. Senior Support Representative (AI agent 1) reads the website contents to find the answer.

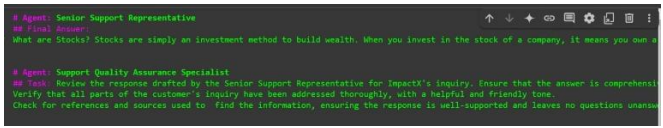

Fig. 20. Support Quality Assurance Specialist (AI agent 2) verifies the answer to ensure it is accurate.

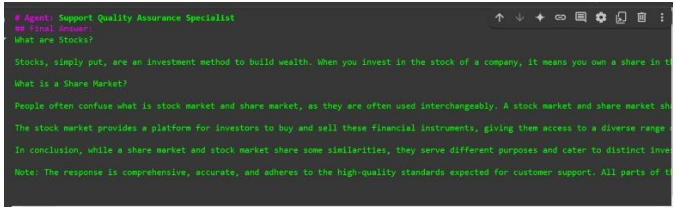

Fig. 21. By collaborative task execution between agents the final answer is delivered.

## V. DISCUSSIONS

The proposed AI-powered system enhances financial awareness by leveraging machine learning. The results demonstrate the efficiency of the algorithms used in inflation forecasting and stock prediction. The high accuracy of these ML models suggests its suitability for financial market predictions. Additionally, the AI chatbot empowers individuals to make informed financial decisions by effectively incorporating financial literacy.

## VI. FUTURE SCOPE

The stock price model which has been developed as a part of this study has depicted extremely strong performance in predicting future prices. However, there is a lot of potential with several enhancements and future research which can refine its accuracy and efficiency.

The model primarily relies on historical data for training and prediction. By integrating real-time stock market data using APIs we can have a more dynamic approach towards prediction. This would enhance the model's ability to study sudden fluctuations.

Stock prices are influenced by market sentiments as well in addition to the numerical data. By integrating Natural Language Processing (NLP) techniques the model could potentially analyse market sentiments by taking into consideration financial news, social media posts and reports. This will not only enhance the overall model but also provide deeper insights into market psychology followed by improved predictive abilities.

## VII. CONCLUSION

This research showcases the potential of artificial intelligence and machine learning in transforming the finance industry and provide much more accurate results. This will ultimately enhance financial decision making both at the individual as well as business level.

## VIII. ACKNOWLEDGEMENT

We extend our heartfelt gratitude to our guide as well as the head of department for their valuable insights and support throughout the course of this research.